\documentclass{ws-p8-50x6-00}
\begin{document}
\title{
\hfill{\normalsize\sf
 Workshop on Strong Magnetic Fields and Neutron Stars,
 Havana, April 6-13, 2003}\\
\vspace{1cm} Interaction of electromagnetic radiation with   supercritical magnetic field}
\author{A. E. Shabad}
\address{P.N.Lebedev Physics Institute, Russian Academy of Sciences
Moscow, Russia
\\ shabad@td.lpi.ru}
\maketitle \abstracts{ It is pointed, that effects of refraction of electromagnetic radiation in the medium, formed by the magnetized vacuum, become essential already for relatively soft photons, not hard enough to create an electron-positron pair, including those belonging to soft gamma-, X-ray, optic and radio- range, if the magnetic field $B$ exceeds the critical value of  $B_{cr}=m^2/e=4.4 10^13$ Gauss.} 
\abstracts{
Three leading terms in the asymptotic expansion of the one-loop polarization operator in a constant magnetic field are found for $B>>B_{cr}$, and the corresponding refraction index is shown to depend only on the propagation direction of the photon relative to the external field. It is established, that  the refraction index for one of polarization modes unlimitedly grows with the field, while the other is saturated at a moderate level. The photon capture effect is extended to soft photons.}
\abstracts{
 The results may be essential
in studying reflection, refraction and splitting of X-rays,
light and radio waves by magnetic fields of magnetars, as well as
in considering emission of such waves by charged particles . }
\newpage
\section{Introduction}
\indent Although it is long since the refracting and birefringing
properties of a strong magnetic field in the vacuum have been
realized, up to now their only essential consequences considered
in a realistic astrophysical context remain the photon splitting
effect \cite{adler} and the effect of photon capture
\cite{nuovocim,annphys,nature,jetp,wunner,ass,bhatia}. The both
effects are currently discussed mostly as applied to electro-magnetic
 radiation
belonging to the $\gamma$-range.  They depend crucially on the
deviation of the photon dispersion curve from the customary shape
it has in the empty vacuum, $k_0^2=|\bf k|^2$, where $k_0$ is the
photon energy, and $\bf k$ is its momentum. As long as one
restricts oneself to considering the magnetic fields $B$ below the
Schwinger critical value $B\leq B_{cr}=(m^2/e)=4.4\cdot 10^{13}$
Gauss, where $m$ and $e$ are the electron mass and charge, resp., the only essential source of this deviation is the singular
behavior of the polarization operator $\Pi_{\mu \nu}(k)$ near the
creation thresholds of mutually independent electron and positron
on Landau levels $n,n'$ by a photon (the cyclotron resonance)
\cite{nuovocim,annphys,nature} or a still stronger singular
behavior of $\Pi_{\mu \nu}$ near the points of a mutually
bound $e^+e^-$-pair (the positronium atom) formation
\cite{jetp,wunner,ass,melrose}. To reach (at least the lower of)
these positionss the photon should belong to the $\gamma$-ray
range, with its energy above or of the order of 1 Mev. For this
reason the effect of photon capture, with its transformation into
electron-positron pair,  derived from  the singular behavior of
$\Pi_{\mu \nu}(k)$, concerns mostly the $\gamma$-quanta, as long
as their propagation in a pulsar magnetosphere of traditional
pulsars is concerned. It was estimated, that the fields about
$B=0.1 B_{cr}$ are sufficient to provide this effect\cite{nature}
and to protect the positronium atom, into which the captured
$\gamma$-quantum is transformed, against ionization by the
accelerating electric field in the polar gap and by the thermal
photons \cite{wunner,jetp,ass,bhatia,melrose}.

 Also the Adler
effect\cite{adler} of photon splitting $\gamma \rightarrow \gamma
\gamma$ for such fields is usually discussed relating to $\gamma$-
quanta \cite{usov,baring,harding,mikheev}. There are two reasons
why, again, the $\gamma$-range is important. The first one is that
the photon splitting becomes possible in the magnetic
 field, because the deviation of the dispersion curve from the $k_0^2=\bf k^2$
 law opens a kinematical aperture for this process - the wider, the stronger
 the deviation ( and the deviation is strong near the thresholds). Besides,
  there is a strong birefringence for the photons in the $\gamma$-range, since
  only one eigenvalue $\kappa_2(k)$ of the tensor $\Pi_{\mu \nu}$ is singular
  near the lowest ($n=n'=0$) threshold, while the other two eigenvalues
  $\kappa_{1,3}(k)$ remain finite, until the next thresholds ($n=0$, $n'=1$
  or $n=1$, $n'=0$) are reached. This implies that the photons of only one
  polarization mode are essentially affected by the medium. This birefringence
  leads to polarization selection rules in the photon splitting process,
  which are well pronounced.
 The second reason is dynamical. The matrix elements of the photon splitting
 are subject to the same resonant behavior near the thresholds as the
 polarization operator. The aforesaid explains why mainly the $\gamma $-range
 is first to be affected by the magnetized vacuum.

 The situation changes considerably after one passes to superSchwinger
 magnetic fields $B\gg B_{cr}$, expected to be existing
in soft gamma-ray repeaters and anomalous X-ray
    pulsars, whose magnetic field is estimated that large ( I refer to the talk,
    given by Prof. A.Reisenegger in this Workshop \cite{reisenegger},
    also to
    \cite{kaspi}). The point is that in such asymptotic regime
a linearly
    growing term, proportional to $B/B_{cr}$, appears \cite{lir,skobelev,mikheev}  in one of the eigenvalues,
    $\kappa_2$, of the polarization operator, thus providing an extra ( apart
    from the cyclotron resonance) large contribution into the refraction of
    the vacuum.

    In Section 3
    we study the   consequences of this phenomenon for the photon
     propagation, basing on the first three leading contributions into
asymptotic expansion of the
     polarization operator eigenvalues for large $B$, obtained
     within the one-loop approximation. One of these consequences
     is a frequency-independent, but direction-sensitive,
    large  refraction index for propagation nonparallel to the magnetic
     field  in one ( out of    three) polarization modes in the kinematical domain far from the threshold.
    The corresponding strong polarization- and direction-sensitive
 refraction occurs for electromagnetic
    radiation of any frequency range, including $X$-ray, optic and radio range.

This study is preceded by Section 2, where  exact results
concerning the electromagnetic radiation propagation in the
magnetized vacuum are described. These follow exceptionally from
the general properties of relativistic, gauge, charge invariance
 \cite{batalin} and Onsager theorem \cite{perez}.
 The results, presented in Section 2, are valid irrespective of any
approximation and the field strength, unless the opposite is
explicitly indicated.

In Appendix the asymptotic expansion, used in Section 3, is
derived.
\section{Exact Facts about Electromagnetic Eigenmodes in an External
Magnetic Field}
 There are three propagating eigenmodes,
corresponding to the vacuum excitations with  photon quantum
numbers in an external magnetic field $\bf$B. The dispersion law,
$i.e.$ the dependence of the energy $k_0$ of the quantum (or the
frequency in the wave) on its momentum $\bf k$, is given for each
mode by a solution of the equation \bea
k^2=\kappa_i(k_0^2-k_\parallel^2,k_\perp^2),\hspace{5mm} &{} &{}
i=1,2,3. \label{6} \eea
 Here $ k_\parallel$ and $ k_\perp$ are the
momentum components along and perpendicular to the magnetic field
$\bf B$, resp., $k^2$ is the photon four-momentum squared,
$k^2=k_\perp^2+k_\parallel^2-k_0^2$. The $\kappa '$s in the
right-hand sides in Eqs.(\ref{6}) are eigenvalues of the
polarization operator.\cite{batalin,nuovocim,annphys}.

It is a general consequence of the relativistic covariance that
the eigenvalues depend upon the two combinations of momentum,
specified in (\ref{6}). This implies, that solutions of the
dispersion equations (\ref{6}) have the following general
structure \bea k_0^2=k_\parallel^2+f_i(k_\perp^2),\hspace{5mm}
i=1,2,3, \label{law} \eea and that the direction of the group
velocity in each mode ${\bf v} = \partial k_0/\partial{\bf k}$
does not coincide (as long as $k_\bot\neq0$) with that of the
phase velocity ${\bf k}/k_0$. To see this, calculate the
components of the group velocity $v_{\bot,\|}$ across and along
the magnetic field $\bf B$, resp., on solutions (\ref{law})  of
each dispersion equation (\ref{6}) \bea\label{group}
  v_\bot\equiv\frac{\partial
  k_0}{\partial k_\bot}=\frac {k_\bot}{k_0}\frac
  {\partial k_0^2}{\partial k_\bot^2}=
  \frac{k_\bot}{k_0}\frac{1-\frac{\partial\kappa_i}{\partial
  k_\bot^2}}{1+\frac{\partial\kappa_i}{\partial
  (k_0^2-k_\|^2)}}=\frac{k_\bot}{k_0}\frac{{\rm d}f_i(k_\bot^2)}{{\rm
  d}k_\bot^2},
\nonumber\\[4pt]
 v_\parallel\equiv\frac{\partial
  k_0}{\partial k_\parallel}=\frac
  {k_\parallel}{k_0}.\hspace{40mm} \eea
  It follows from (\ref{group}) that the angle
$\theta$ that the direction $\bf v$ of the electromagnetic energy
propagation makes with the external magnetic field satisfies the
relation \be \frac{v_\perp}{v_\|}\equiv\tan\theta
=\left(1-\frac{\partial\kappa_i}{\partial
  k_\bot^2}\right)
  \left(1+\frac{\partial\kappa_i}{\partial
  (k_0^2-k_\|^2)}\right)^{-1}\tan\vartheta, \label{11} \ee where $\vartheta$ is the angle
between the photon momentum (phase velocity) and the external
field, $\tan\vartheta\equiv k_\perp/k_\|$. The following statement
takes place: if the phase velocity exceeds the velocity of light
$c$, $i.e.$ if  $k_\bot^2+k_\|^2>k_0^2$, (or
$f_i(k_\bot^2)<k_\bot^2$ in (\ref{law}) ), but the group velocity
(\ref{group}) does not, $v_\bot^2+v_\|^2\leq1$~ (or d$^2f_i(k_\perp^2
)/({\rm d}k_\perp^2)^2<0$, then
$\tan\theta<\tan\vartheta$. The conditions of this statement are
fulfilled for the dispersions laws found within the
approximation-dependent calculations of the $\kappa$'s. For the
superSchwinger  fields, treated within one-loop approximation,
this fact follows explicitly from equations of Section 3 below.
Therefore, the photon tends to deviate closer to the magnetic
field line, as compared to its momentum direction.

It follows from the gauge invariance that \bea \kappa_i(0,0)=0,
\hspace{5mm} i=1,2,3. \label{gauge} \eea This property implies
that for each mode there always exists a dispersion curve with
$f_i(0)=0$, which passes through the origin in the
$(k_0^2-k_\parallel^2, k_\perp^2)$-plane. However, out of these
three solutions only two may simultaneously correspond to physical
massless particles - the photons. The third one is a nonphysical
degree of freedom, characteristic of gauge theories: in a magnetic
field a photon has two degrees of freedom, the same as in the
empty vacuum. Which of the modes becomes nonphysical, depends upon
direction of propagation and specific form of the function
$f_i(k_\perp^2)$ in (\ref{law}). We shall further discuss this
matter for the  superSchwinger field limit in the next Section.
Massive branches of solutions of (\ref{6}), with $f_i(0)>0$, may
also exist, despite (\ref{gauge}). For them the number of physical
degrees of freedom is three, so all the three equations (\ref{6})
may have physical solutions simultaneously (see, $e.g.$ the
positronium branches found in \cite{ass,trudy,book})

The refraction index $n_i$ in Mode i is \bea n_i\equiv\frac{|{\bf
k}|}{k_0}= \left( 1+\frac{\kappa_i}{k_0^2}\right)^{\frac 1{2}}
=\left(1+\frac{k_\bot^2-f_i(k_\bot^2)}{k_0^2}\right)^\frac 1{2}.
\label{index} \eea Unlike $\kappa_i$, the refraction index $n_i$ is not a Lorentz scalar and may depend on two energy-momentum variables, after it is reduced onto 
the dispersion law (\ref{law}). The gauge invariance property (\ref{gauge})
implies that the refraction index (\ref{index}) for parallel
propagation, $k_\bot=0$, be exactly equal to unity for the
massless, $f_i(0)=0$, branches in every mode
\begin{equation}\label{npar}
  n_i^\|=1.
\end{equation}
The electromagnetic wave propagating strictly along the external
constant and homogeneous magnetic field does this with the
velocity of light in the vacuum $c$, the phase and group
velocities coinciding in this case.

If, within a certain approximation, the eigenvalue $\kappa_i$ is a
linear function of its arguments with the condition (\ref{gauge})
met, the refraction index (\ref{index}) for the corresponding dispersion law
depends on a single combination of the photon energy and momentum,
which is the direction of propagation $\vartheta$. This happens in
a nonresonant situation, for instance, as described in the next
section.

 The polarizations of the modes are described in an
approximation- independent way\cite{batalin,annphys} by the
following relations, where $\bf e^{(i)}$ and $\bf h^{(i)}$ are the
electric and magnetic fields in the wave belonging to Mode number
$i=1,2,3$ \bea {\bf e}^{(1)} =-\frac{{\bf k_\perp}}{k_\perp}k_0,
\hspace{35mm} {\bf h}^{(1)} =
(\frac{\bf k_\perp}{k_\perp} \times {\bf k}_\parallel),  \\
{\bf e}^{(2)}_\perp = {\bf k}_\perp k_\parallel, \hspace{7mm}{\bf
e}^{(2)}_\parallel = \frac{{\bf k}_\parallel}{k_\parallel}
(k_\parallel^2-k_0^2),\hspace{3mm}     {\bf h}^{(2)} =
-k_0\left({\bf k_\perp}\times \frac{{\bf
k_\parallel}}{k_\parallel}\right),
\\
{\bf e}^{(3)} = - k_0 \left(\frac{{\bf
k_\perp}}{k_\perp}\times\frac{{\bf
k_\parallel}}{k_\parallel}\right),\hspace{3mm}{\bf
h_\perp}^{(3)}=-\frac{\bf k_\perp}{k_\perp}
k_\parallel,\hspace{3mm} {\bf h_\parallel}^{(3)}= \frac{\bf
k_\parallel}{k}_\parallel k_\perp. \label{5} \eea Here the cross
stands for the vector product, and the boldfaced letters with
subscripts $\parallel$ and $\perp$ denote vectors along the
directions, parallel and perpendicular to the external magnetic
field, resp. The electric field $\bf e$ in the  wave of Mode 1 is
parallel to $\bf k_\perp$, in Mode 2 it lies in the plane
containing the vectors $\bf k, B$, in Mode 3 it is orthogonal to
this plane, $i.e.$ Mode 3 is always transversely polarized.

Note, that the normalization is different for each line of
(\ref{5}), so one may judge about vanishing of some components as
compared to others within one line, but not between different
lines.

Concerning the direction of propagation, two cases are essentially
different. If $k_\perp=0$, we speak about longitudinal
propagation. If not, a Lorentz boost along the external (constant
and homogeneous) magnetic field exists, which does not  change the
value of the latter and does not introduce an extra electric
field, but nullifies $k_\parallel$. Hence, the general case of
nonparallel propagation $k_\perp\neq0$, $k_\parallel\neq 0$ is
reduced to purely transversal propagation, $k_\parallel=0$ (in the
corresponding reference frame). One should keep in mind, however,
that the above transformation changes the photon energy $k_0$ and
should be treated with precautions when one considers a field with
curved lines of force.

 For transversal propagation,
 ${\bf k}\perp{\bf B}$, $(k_\parallel=0)$,  Modes 2, 3 are transversely
 polarized
 ${\bf e}^{(2,3)}\perp{\bf k}$ in two mutually orthogonal planes,
${\bf e}^{(2)}\perp{\bf e}^{(3)}$, while Mode 1 is longitudinally
polarized,
 ${\bf e}^{(1)}\parallel{\bf k}$ with no magnetic field in it,
${\bf h}^{(1)}=0$. It is expected not to correspond to a photon
(depending on the dispersion law).

On the contrary, for longitudinal propagation, ${\bf k}\parallel
{\bf B}$, $(k_\perp=0),$ Modes 1, 3 are transversely polarized,
${\bf e}^{(1,3)}\perp{\bf B}$, and their electric field vectors
lie in mutually orthogonal planes, ${\bf e}^{(1)}\perp{\bf
e}^{(3)}$, as they always do, while Mode 2 is longitudinally
polarized, ${\bf e}^{(2)}\parallel{\bf B}$, and does not contain a
magnetic field, ${\bf h}^{(2)}=0$. This time Mode 2 is expected
not to correspond to a photon, whereas Mode 1 is a physical
electromagnetic wave, which matches the electromagnetic wave of
Mode 3: together they may form a circularly polarized transversal
wave because of the degeneracy property \be
\kappa_1((k_0^2-k_\parallel^2),0)=
\kappa_3((k_0^2-k_\parallel^2),0). \label{degeneracy} \ee The
latter reflects the cylindrical symmetry of the problem of a
photon propagating along the external magnetic field.

Another remark of almost general character is in order. One might
expect the possibility of the Cherenkov radiation by a charged
particle moving in an optically dense medium formed by the
magnetized vacuum. 
This effect - with the Cherenkov photons softer than $k_0=2m$ -
does not take place in known situations, however. Consider an emission of a
photon by an electron in a magnetic field, not accompanied by the
change of its Landau quantum number, $n=n'$ (otherwise, that would
be the cyclotron, and not Cherenkov, radiation).  According to the
kinematical analysis of the energy and momentum conservation, done
in\cite{2perez}, (also to the study\cite{2perez} of analyticity
regions of the one-loop photon polarization  operator in
electron-positron plasma in a magnetic field, calculated
in\cite{perez}), the Cherenkov photon with $k_0<2m$ may only belong to the right
lower sector \be k_0^2-k_\parallel^2\leq 0,
\hspace{3mm}k_\perp^2\geq 0 \label{cherenkov} \ee in the
$(k_0^2-k_\parallel^2,k_\perp^2)$-plane. The substantial reason
for this circumstance is the degeneration of the electron energy
with respect to the center-of-orbit position in the transversal
plane. No dynamical calculations, hitherto known, provide a
penetration of photon dispersion curves into this sector. The only
exception is a nonphysical
 situation  due to exponentially strong external fields, to be mentioned
in Subsection 3.2 below. We conclude that under standard
conditions no Cherenkov emission of a photon softer than $2m$ is possible.
\section{Photon Dispersion in SuperSchwinger Magnetic Field}
\subsection{Asymptotic expansion  of polarization tensor eigenvalues}
In the asymptotic region of supercritical magnetic fields $B\gg
B_{cr}$ and restricted energy of longitudinal motion
$k_0^2-k_\|^2\ll (B/B_{cr})m^2$, the three eigenvalues
$\kappa_{1,2,3}(k)$ of the polarization operator - provided the
latter is calculated within the one-loop approximation as done in
\cite{batalin,tsai} - have the following behavior, derived from
equations of Ref.\cite{annphys} (see Appendix) \bea
\kappa_1(k_0^2-k_\parallel ^2,k_\perp^2)=\frac{\alpha k^2}
{3\pi}\left(\ln \frac{B}{B_{cr}}-C-1.21\right), 
\label{1} \eea \bea
\kappa_2(k_0^2-k_\parallel^2,k_\perp^2)=\hspace{80mm}\nonumber \\[4pt]
=\frac{\alpha Bm^2(k_0^2-k_\|^2)}{\pi B_{cr}}\exp \left(-
\frac{k_\bot^2}{2m^2}\frac{B_{cr}}{B}
\right)\int_{-1}^1\frac{(1-\eta^2)\rm d
\eta}{4m^2-(k_0^2-k_\|^2)(1-\eta^2)},\nonumber\\[4pt] \label{2}\eea
\bea \kappa_3(k_0^2-k_\parallel^2,k_\perp^2)=\hspace{80mm}\nonumber \\[4pt]
=\frac{\alpha k^2} {3\pi}\left(\ln \frac{B}{B_{cr}}-C \right)
-\frac
\alpha{3\pi}\left(0.21k_\perp^2-1.21(k_0^2-k_\parallel^2)\right).\hspace{20mm}
\label{kappa3} \eea Here $\alpha=1/137$ is the fine structure
constant, $C=0.577$ is the Euler constant.
 Eqs.~(\ref{1}),~(\ref{kappa3}) are accurate up to
terms, decreasing with $B$ like $(B_{cr}/B)\ln (B/B_{cr})$ and
faster. Eq.~(\ref{2}) is accurate up to terms, logarithmically
growing with $B$. In $\kappa_{1,3}$ we took also the limit
$k_\perp^2\ll(B/B_{cr})m^2$, which is not the case for $\kappa_2$,
wherein the factor $~~\exp~ (-k_\perp^2B_{cr}/2m^2B)~~$ is kept
different from unity, because it is important near the cyclotron
resonance, as it will be explained in Subsection 3.2 below. The
integral in (\ref{2}) can be readily calculated, but we shall not
need its explicit form here.

The parts growing with $B$ in $\kappa_{1,2,3}$ 
were written in \cite{lir}, their derivation from equations of
Ref. \cite{annphys} is traced in detail in \cite{trudy,book}. The
linearly growing term in Eq.~(\ref{2}) in a different way was
obtained in \cite{skobelev} using a two-dimensional (one time, one
space) diagrammatic technique developed  to serve the asymptotic
magnetic field regime. The logarithmic terms in the expressions
above do not dominate over the constant terms unless one would
like to include
 exponentially large magnetic fields into
consideration. (That would be unreasonable not only because such
fields are hardly expected to exist in nature, but mainly because
their consideration is beyond the scope of Quantum
Electrodynamics: the logarithmically growing terms in ~(\ref{1}),
 (\ref{kappa3}) are associated with the absence of
asymptotic freedom in QED ($cf$. analogous asymptotic behavior
\cite{ritus2} in the Euler-Heisenberg effective Lagrangian).) The
derivation of all  terms in Eqs.~(\ref{1}), (\ref{2}),
(\ref{kappa3}), including those that do not grow with $B$, is
given in Appendix using a straightforward method different from
the one applied earlier in \cite{trudy,book}. The asymptotic
expressions used in \cite{mikheev} do not coincide with ours,
except for the linear-in-B term.

The limiting expressions (\ref{1}), (\ref{2}), (\ref{kappa3}) do
satisfy the exact properties (\ref{degeneracy}) and (\ref{gauge}).

 In this paper we shall only deal with
the transparency  region, $k_0^2-k_\|^2\leq 4m^2$ ($i.e.$ with the
kinematical domain, where $\kappa_{1,2,3}$ are real), because we
shall be interested in photons with $k_0<2m$, even $k_0\ll 2m$,
which never reach the free pair creation threshold
$k_0^2-k_\|^2=4m^2$.  The eigenvalue $\kappa_2$ (\ref{2}) has a
singular branching point in the complex plane of the variable
$(k_0^2-k_\|^2)$ near the lowest pair creation threshold
$(k_0^2-k_\|^2)_{thr}=4m^2$. Thresholds of creation of
$e^+e^-$-pairs with the electron and the positron on excited
Landau levels $ n,n'\neq 0 $ \be\label{thresh}
(k_0^2-k_\|^2)_{thr}^{n,n'} =m^2\left[\left(
1+n\frac{B}{B_{cr}}\right)^\frac{1}{2} +\left(
1+n'\frac{B}{B_{cr}}\right)^\frac{1}{2}\right]^2 \ee are shifted
in the asymptotic regime to infinitely remote region. For this
reason the eigenvalues $\kappa_{1,3}$, which are responsible for
photons of such polarizations that can only create $e_+e_-$-
 pairs with at least one charged particle in an excited Landau
 state,
 do not contain imaginary parts or singular branching
  points in this regime. On the other hand, the eigenvalue $\kappa_2$ has only one
  singular branching point, corresponding to possibility of creation of
  electron and positron in the lowest Landau states by the photon, polarized as
  in Mode 2.
  The singular threshold behavior of (\ref{2}) near the point
  $k_0^2-k_\|^2=4m^2-\epsilon$, $\epsilon>0$, $\epsilon\rightarrow 0$ is
\bea \kappa_2(k)\sim \frac{2\alpha
Bm^3}{B_{cr}}\exp\left(-\frac{k_\perp^2}{2m^2}
\frac{B_{cr}}{B}\right)\left(
4m^2-k_0^2+k_\parallel^2\right)^{-\frac{1}{2}}. \label{4} \eea As
might be expected, this is the same as the behavior near this
threshold of the exact one-loop expression for $\kappa_2(k)$
\cite{annphys}, before the limiting transition to large fields has
been performed.

\subsection{Propagation of eigenmodes in the
superSchwinger field limit}

If Eq.(\ref{1}) for $\kappa_1$ is taken as the r.-h. side of
Eq.~(\ref{6}), the latter has only one  solution, which is the
trivial dispersion law $k^2=0$. With the relation $k^2=0$
satisfied, however, the 4-potential corresponding to the
electromagnetic field of Mode 1 becomes proportional to  the
photon 4-momentum vector $k_\mu$, unless $k_\|=0$ (see
\cite{annphys,trudy,book}). Therefore, for nonparallel
propagation, Mode 1 corresponds only the gauge degree of freedom
we discussed in Section 2, with no real electromagnetic field
associated with it.

 Solutions of equation (\ref{6})  for the second mode $i=2$  with Eq.(\ref{2})
taken for $\kappa_2$ are plotted in the Figure for three values of the field $B$ using MATHCAD code. These are  
 dominated by the cyclotron
 resonance (\ref {4}), that causes the strong deviation of
the dispersion curves in Fig.1 from the shape $k^2=0$ ( the light
cone ). Once $k_\perp^2\rightarrow\infty$ near the threshold on
the
 dispersion curves, the
quantity $~k_\perp^2B_{cr}/m^2B$ should be kept different from
zero even in the large field limit under consideration.
\begin{figure}[t]
\epsfxsize=25pc 
\epsfbox{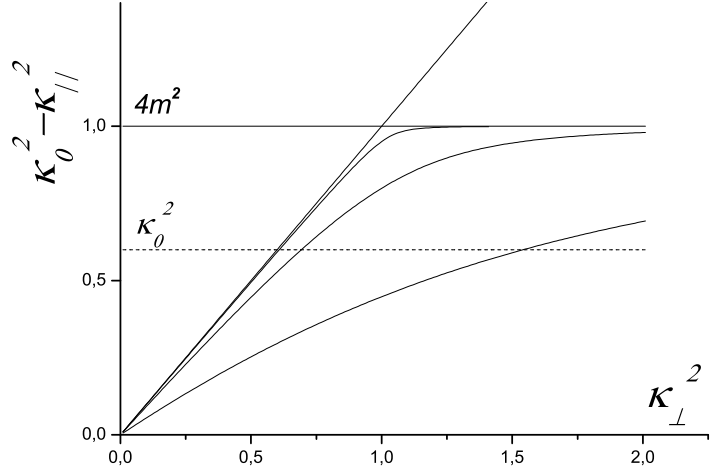} 
\caption{A family of dispersion curves for Mode 2 - solutions of
equation (\ref{6}), with Eq.(\ref{2}) taken for the r.-h. side, -
below the threshold $k_0^2-k_\parallel^2=4m^2$. The values of the
external magnetic field, corresponding to the curves, are ( from
left to right) $B=10 B_{cr}, 100 B_{cr}, 1000 B_{cr}$. The
straight line is the light cone - dispersion curve for $B=0$. The
dashed horizontal line marks the maximum to which the photon with
the energy $k_0$ may proceed , provided that $k_0<2m$. The
variables along the axes are plotted in units of $4m^2\approx 1
Mev^2$} \label{fig} \vspace{-0.5cm}
\end{figure}

The behavior of the dispersion curves  of Mode 2 near the
threshold for superSchwinger magnetic fields $B\gg B_{cr}$ is the
same as for the "moderate" fields $B\leq B_{cr}$, and therefore it
also presents the photon capture effect for photons harder than
$2m$, known for such fields\cite{nature}: if one calculates
(\ref{11}) near the threshold $k_0^2-k_\parallel^2=4m^2$ using Eq.
(\ref{4}) as $\kappa_2$ to get
\begin{equation}\label{capture}
  \tan\theta=\frac{k_\bot}{k_\|}\frac{B_{cr}}{Bm^2}(4m^2-k_0^2+k_\|^2),
\end{equation}
one concludes that the angle $\theta$ between the external
magnetic field and the direction of the wave packet propagation in
Mode 2 tends to zero, the faster the stronger the field. If the
photon energy $k_0$ is just a little less than $2m$, the photon
may be close to the threshold when its $k_\|$ disappears. In this
upper point the wave packet stops, since the group velocity length
$v_\bot^2+v_\|^2=v_\|^2(1+\tan^2\theta$), equal to $k_\|^2/k_0^2$
according to the second line of (\ref{group}) and (\ref{capture}),
disappears together with $k_\|$.

As applied to the conventional pattern of a pulsar magnetosphere,
already described in the talk by Prof. Qiao\cite{qiao}, this
effect acts as following\cite{nature}. A curvature $\gamma
$-quantum emitted tangentially to the magnetic line of force,
$i.e.$ placed initially in the origin of Fig.1, then evolves along
its dispersion curve as it propagates in the dipole magnetic field
with its line of force curved, since the components $k_\parallel$
and $k_\perp$ are changing. The maximum value of the ordinate
$k_0^2-k_\parallel^2$ occurs at $k_\parallel=0$, and  it is the
photon energy squared, $k_0^2$. If the latter is greater than
$4m^2$, the photon may achieve the horizontal asymptote in Fig.1.
Here its group velocity across the magnetic field,
d$k_0/$d$k_\perp$, disappears, d$k_0/$d$k_\perp\rightarrow 0$, and
hence it propagates along the magnetic  field and does not cross
the threshold, since the other branch of the dispersion curve,
which passes above the threshold, is separated by a gap from the
initial branch. As a matter of fact, a mixed state: photon-pair is
formed\cite{nature}, analogous to polariton known in condensed
matter physics. The massless part of its spectrum is presented by
the dispersion curves in Fig.1. The photon gradually turns into
the $e^+-e^-$-pair and exists mostly in that form when
 finally it is propagating along the magnetic lines of force. This
capturing effect is important for the formation of radiation of
pulsars with the fields $B>0.1 B_{cr}$, since it prevents - once
the binding of the electron-positron pair into a positronium atom
is taken into account\cite{jetp,wunner,ass,bhatia,melrose} - the
screening of the accelerating electric field in the polar gap. It
may be essential for magnetars with their fields $\sim 10^{14} -
10^{15}$ Gauss, as well.

The new features, introduced by superSchwinger fields are that the
dispersion curves for Mode 2, presented in Fig.1, step aside from
the light cone already far from the resonance region. This means
that although the photons softer than $2m=1 Mev$ cannot proceed to
the values of the ordinate in Fig.1 higher than their energy
squared (corresponding to $k_\|=0$), they can still reach the
region where the transversal group velocity d$k_0/$d$k_\perp$
becomes much less than unity and are thus captured to the
trajectory almost parallel to the magnetic field. This is how the
capture effect
 extends to the photon energies below the border $k_0=2m$. The
 cyclotron singularity at the pair-creation threshold in such fields
 is so strong that  even  low energy photons that are unable
 to create a pair are sensitive to it, provided that they belong to Mode 2!

 Besides the extension of the photon capture effect to softer photons, the
 inclusion of superSchwinger fields into consideration has another impact.
 It leads to large direction-dependent refraction of  Mode 2 electromagnetic
 waves of low frequency. To see this, consider the limit
 \be
 k_0^2-k_\|^2\ll4m^2
 \label{7}
 \ee
in Eq.(\ref{2}),  which reduces to neglecting $k_0^2-k_\|^2$ in
the integrand in (\ref{2}). Then (\ref{2}) becomes \be
\kappa_2(k)=\frac{\alpha}{3\pi} (k_0^2-k_\|^2)\frac{B}{B_{cr}}
\exp\left(-\frac{k_\perp^2}{2m^2}\frac{B_{cr}}{B}\right).
\label{7a} \ee The exponential factor in (\ref{7a}) cannot be
essential within the region (\ref{7}). The dispersion equation
(\ref{6}) for Mode 2 ($i=2$) then has the solutions expressing the
photon energy $k_0$ as a function of its transversal and
longitudinal momentum:  \be
k_0^2=k_\|^2+k_\perp^2\left(1+\frac\alpha{3\pi}\frac{B}{B_{cr}}\right)^{-1}.
\label{9} \ee Eq.(\ref{9}) presents analytically the straight line
parts of the dispersion curves in Fig.1 adjacent to the origin for
various values of $B$. The components of the group velocity
(\ref{group}) $ v_{\perp,\|}$ calculated from (\ref{9}) are \bea
v_\perp&=&\frac{k_\perp}{k_0} \left(1+\frac\alpha{3\pi}\frac
B{B_{cr}}\right)^{-1},\hspace{5mm}
 v_\|=\frac{k_\|}{k_0}. \label{10} \eea
The modulus of the group velocity squared is now
\begin{equation}\label{modulus}
  v_\bot^2+v_\|^2=\frac 1{1+\frac\alpha{3\pi}\frac
B{B_{cr}}}+\frac{\frac\alpha{3\pi}\frac
B{B_{cr}}\cos^2\vartheta}{1+\frac\alpha{3\pi}\frac
B{B_{cr}}\cos^2\vartheta},
\end{equation}where
$\vartheta$ is the angle between the photon momentum and the
field, $\tan\vartheta=k_\perp/k_\|$. Eq.(\ref{modulus}) has the
maximum value of unity for the parallel propagation,
$\vartheta=0$,  in accordance with the general statement of
Section 2, and is minimum for perpendicular propagation,
$\vartheta=\pi/2$.

The expression (\ref{11}) for the angle $\theta$ between the
direction of the electromagnetic energy propagation and the
external magnetic field in the superSchwinger limit for Mode 2
becomes \be \frac{v_\perp}{v_\|}=\tan\theta=
\frac{k_\perp}{k_\|}\left( 1+\frac\alpha{3\pi}\frac
B{B_{cr}}\right)^{-1} =\tan\vartheta\left(
1+\frac\alpha{3\pi}\frac B{B_{cr}}\right)^{-1}. \label{} \ee Once
$\tan\theta<\tan\vartheta$, in the problem, formulated above, the
photon emitted tangentially to curved lines of force  will bend
towards these lines. This relates to low-frequency radiation, as
well.

The refraction index (\ref{index}) in Mode 2 for
$k_0^2-k_\|^2\ll4m^2$, $B\gg B_{cr}$ is \bea n_2
=\left(\frac{1+\frac\alpha{3\pi}\frac
B{B_{cr}}}{1+\frac\alpha{3\pi}\frac
B{B_{cr}}\cos^2\vartheta}\right)^\frac 1{2} . \label{12} \eea The
refraction index obtained depends upon the direction of the photon
momentum, characterized by the angle $\vartheta$, but does not
depend upon its energy. In other words, there is no frequency
dispersion in a wide range from slow radio waves and up to soft
$\gamma$-rays with $k_0\ll 2m$. This is a consequence of the fact,
that in $\kappa_2$
 only linear parts in momenta squared were as a
matter of fact left ( correspondingly, $f(k_\bot^2)$ in
(\ref{law}) is proportional to $k_\bot^2$ according to (\ref{9})).

 The refraction index (\ref{12}) reaches its maximum for
transversal propagation ($k_\|=0$, $\vartheta=\pi/2$) \be
n_2^\bot=\left(1+\frac\alpha{3\pi}\frac B{B_{cr}}\right)^\frac
1{2}=\left(1+7.7\cdot 10^{-4}\frac B{B_{cr}}\right)^{\frac 1{2}}.
\label{13} \ee  For $B\sim 10 \cdot B_{cr}$ the declination of the
refraction index (\ref{13}) from unity exceeds that value for
gases at atmospheric pressure in optic range by an order of
magnitude, for $B\sim 1000\cdot B_{cr}$ it reaches the value,
characteristic of transparent liquids and glass; the refraction
index (\ref{13}) becomes equal to that of diamond ($n=2.4$) for
$B=27\cdot 10^{16}$Gauss.

Contrary to the just considered case of Mode 2, the polarization
tensor eigenvalue $\kappa_3$ (\ref{kappa3}) does not contain a
contribution linearly growing with the external field, as well as
the resonance. For Mode 3 the dispersion equation (\ref{6}) with
its r.-h. side given as (\ref{kappa3}) has the solution
\begin{equation}\label{dis3}
  k_0^2=k_\|^2+k_\bot^2\frac{Z-\frac\alpha{3\pi}}{Z},
\end{equation}
where
\begin{equation}\label{Z}
  Z =1-\frac\alpha{3\pi}\left(\ln\frac B{B_{cr}}-C-1.21\right).
\end{equation}

The known absence of asymptotic freedom in QED manifests itself in
the negative sign in front of the logarithm in (\ref{Z}). This
results in pathological consequences for the fields, as large as
$\sim B_{cr}\exp(3\pi/\alpha)$. In this domain the coefficient of
$k_\perp^2$ in (\ref{dis3}) becomes, as the field grows, first
less than zero and later greater than unity. The corresponding
dispersion laws are nonphysical, since they lead to the group
velocity greater than unity.  In the negative slope case in
(\ref{dis3}), $e>B\exp (-0.21- C -3\pi/\alpha)/B_{cr}>1$, the
dispersion curve enters the sector (\ref{cherenkov}) acceptable
for the Cherenkov radiation. But this is the Cherenkov emission of
tachyons!  It is also odd, that in the latter case  e.-m. waves
may only propagate inside the cone $0<\tan
\vartheta<-1+\alpha/3\pi Z$ with its axis along the external
field, irrespective of the way they are produced. This domain of
exponentially large external fields is not of our interest in the
present paper.

For the fields that are not exponentially large, with the
logarithmic terms of the order of unity, one should treat all the
terms, marked by the coefficient $\alpha/3\pi$ in (\ref{dis3}), as
small. Then, finally, the dispersion law for Mode 3 becomes:
\begin{equation}\label{fin3}
  k_0^2=k_\|^2+k_\bot^2\left(1-\frac\alpha{3\pi}\right)
\end{equation}Notably, the  field-containing logarithmic terms have cancelled
from here. Thus the dispersion law (\ref{fin3}) of Mode 3 is
saturated in the sense that, unlike Eq.(\ref{9}) for Mode 2, it
has reached the universal form, independent of the external field
in the superSchwinger limit. The refraction index of Mode 3
corresponding to (\ref{fin3}) is
\begin{equation}\label{n3_}
  n_3=1+\frac\alpha{6\pi}\sin^2\vartheta.
\end{equation}
Again, the same as (\ref{13}), the maximum refraction in Mode 3 is
achieved at perpendicular propagation, $\vartheta=\pi/2$: \be
n_3^\bot=1+3.8\cdot 10^{-4}. \label{perp3} \ee This refraction
index is of the order of that of gaseous ammonia and cannot be
made larger by increasing the external field any further.
\section{Conclusion}
We have found that in the asymptotic case of external magnetic
fields $B$ that can be orders of magnitude larger than the
Schwinger value of $4.4\times 10^{13}$Gauss, the refractive
capacity of the magnetized vacuum grows unlimitedly with this
field for electromagnetic radiation belonging to the polarization
Mode 2, but comes to a saturation at a moderate level of
corrections $\sim \alpha/3\pi$ for Mode 3. For the "parallel
energy" of the photon not close to the cyclotron resonance,
$k_0^2-k_\|^2\ll 4m^2$, the refraction effects for Mode 2
essentially exceed the  above small corrections, typical for
nonasymptotic domain, already for $B\sim 10\cdot B{cr}$. A regime
is established in the range of photon frequencies/energies,
extending from zero and up to soft $\gamma$-rays, for which the
dispersive properties of the magnetized vacuum are in each mode
independent of the photon frequency/energy, but do depend upon the
direction of its propagation. Apart from the fact that the
refraction index in Mode 2 for the propagation nonparallel to the
external field numerically grows with the field, it is remarkable
that the angle between the group velocity and the direction of the
photon momentum grows, too, the wave packet being attracted by the
line of force of the external field. In this way the effect of
$\gamma$-quantum capture by a strong magnetic field, known to
exist due to resonance phenomena associated with free and
bound pair creation, is extended to lower energy ranges.
Therefore, not only hard $\gamma$-rays, but also X-rays, light and
radio-waves undergo strong dispersive influence of the magnetized
vacuum, when the magnetic fields are of the order of magnitude of
those estimated to exist in magnetars. In view of this fact the
e-m energy canalization phenomena may become important not only
within the traditional context described in Subsection 3.2 above,
but also as applied to the scattering of e-m waves falling onto
the magnetic field from outside\cite{nuovocim}. These may be, for
instance, the X-rays emitted from the accretion disk or from the
pulsar surface outside the region where the magnetic field enters
it. Recently the problem of the bending of electromagnetic
radiation by the dipole magnetic field of a neutron star was
addressed\cite{denisov}, and the competition of this process with
the effects of gravity was considered. (The author is indebted to
Prof. H. Mosquera Cuesta who attracted his attention to that work
in the course of Workshop). We insist, however, that such effects
 cannot
be adequately treated disregarding the refraction index dependence
on the direction of propagation and using the
quadratic-in-the-field expressions for polarization operator, only valid in the low
field limit, as is the case in Ref.\cite{denisov}.
\section*{Acknowledgements}
I am indebted to Professor Hugo P$\acute{e}$rez Rojas for the
hospitality, extended to me during the Workshop on Strong Magnetic
Fields and Neutron Stars at ICIMAF in Havana,  and for encouraging
me to refresh the study of magnetic optics of the vacuum. I
acknowledge the financial support  of Russian Foundation for
Fundamental Research ( RFFI 02-02-16944 ) and the President of
Russia Program for Support of Leading Scientific Schools
(LSS-1578.2003.2). My stay in Havana became possible thanks to the
support granted by Instituto de Cibern$\acute{e}$tica,
Matem$\acute{a}$tica y F$\acute{i}$sica (ICIMAF), Centro
Latino-Americano de F$\acute{i}$sica (CLAF) and Abdus Salam
International Centre for Theoretical Physics (ICTP).
\section*{Appendix}
In this Appendix we derive the asymptotic expansion, presented in
Subsection  3.1, from expressions of
Ref.\cite{annphys,trudy,book}.

The three eigenvalues $\kappa_i$, $i=1,2,3$ of the photon
polarization operator in the one-loop approximation, calculated
using exact propagator of electron in an external magnetic field,
can be expressed as linear combinations of the three functions
$\Sigma_i$ \bea\label{ro123}
 \kappa_1=-\frac 1{2}\left(z_1+z_2\right)\Sigma_1,\hspace{5mm}\kappa_2=
  -\frac 1{2}\left(z_1\Sigma_2 +z_2\Sigma_1 \right), \nonumber\\[4pt]
  \kappa_3=-\frac 1{2}\left(z_2\Sigma_3 +z_1\Sigma_1 \right),
\eea where  the new notations for the momentum variables:
\begin{equation}\label{momenta}
  z_1=k_\|^2-k_0^2,\hspace{5mm}z_2=k_\bot^2
\end{equation} are introduced, $k^2=z_1+z_2$.
Here $\Sigma_i$ are dimensionless functions of the three ratios
$B_{cr}/B,~~z_2 B_{cr}/m^2 B,
  ~~z_1 B_{cr}/m^2 B,~~$ equal to
\begin{equation}\label{Sigma}
  \Sigma_i=\Sigma_i^{(1)}+\Sigma_i^{(2)},
\end{equation}
\bea\label{A}
  \Sigma_i^{(1)}\left(\frac{B_{cr}}{B}\right)=\frac{2\alpha}{\pi}\int_0^\infty{\rm d}t~
  \exp\left(-\frac{tB_{cr}}{B}\right)\int_{-1}^1{\rm
  d}\eta\left[\frac{\sigma_i(t,\eta)}{\sinh t}-\lim_{t\rightarrow 0}
  \frac{\sigma_i(t,\eta)}{\sinh
  t}\right],\nonumber\\[4pt]
\eea

 \bea\label{B}
  \Sigma_i^{(2)}\left(\frac{B_{cr}}{B},\frac{z_2 B_{cr}}{m^2 B},
  \frac{z_1 B_{cr}}{m^2 B}\right)=
  \frac{2\alpha}{\pi}\int_0^\infty{\rm d}t~
  \exp\left(-\frac{tB_{cr}}{B}\right)\int_{-1}^1{\rm
  d}\eta\frac{\sigma_i(t,\eta)}{\sinh t}\times\nonumber\\[4pt]
  \times\left[\exp\left(-z_2\frac{M(t,\eta)}{eB}-z_1
  \frac{1-\eta^2}{4eB}t\right)-1\right],
\nonumber\\[4pt]
\eea where
\begin{equation}\label{M}
  M(t,\eta)=\frac{\cosh t-\cosh t\eta}{2\sinh t},
\end{equation}
and
\begin{equation}\label{sigma1}
  \sigma_1(t,\eta)=\frac{1-\eta}{2}~\frac{\sinh(1+\eta)t}{2\sinh t},
\end{equation}
\begin{equation}\label{sigma2}
\sigma_2(t,\eta)=\frac{1-\eta^2}{4}\cosh t,
\end{equation}
\begin{equation}\label{sigma3}
\sigma_3(t,\eta)=\frac{\cosh t-\cosh\eta t}{2\sinh^2 t}.
\end{equation}
The designation "lim" in (\ref{A}) stands for the asymptotic limit
\begin{equation}\label{limit}
  \lim_{t\rightarrow 0}\frac{\sigma_i(t,\eta)}{\sinh
  t}=\frac{1-\eta^2}{4t},
  \hspace{5mm}i=1,2,3.
\end{equation}
The fact that $\Sigma_i$ do not depend on the fourth possible
dimensionless variable $z_1/z_2$ ~~seems to be an
approximation-independent manifestation of analyticity properties
due to dispersion relations of Kramers-Kronig nature.

 Consider
$\Sigma_i^{(1)}$ first. It does not depend on the photon energy
and momentum. With the use of the notation
\begin{equation}\label{g's}
  g_i(t)=\int_{-1}^{1}\sigma_i(t,\eta)~{\rm d}\eta,
\end{equation}
(\ref{A}) can be represented as \bea\label{Sigma(1)}
\Sigma_i^{(1)}=\frac{2\alpha}{\pi}\int_0^\infty\exp\left(-\frac{tB_{cr}}{B}\right)
  \left(
  \frac{g_i(t)}{\sinh t}-\frac
  1{3t}\right){\rm d}t
\eea
 The integrals ( \ref{g's}) are explicitly
calculated to give \bea\label{g}
  g_1(t)=\frac 1{4t\sinh t}\left(\frac{\sinh
  2t}{t}-2\right),\hspace{5mm} g_2(t)=\frac{\cosh t}{3},\nonumber\\[4pt]  g_3(t)=\frac 1{\sinh^2t}\left(\cosh t-\frac{\sinh t}{t}\right).\hspace{25mm}
\eea
 Our goal now is to find the asymptotic behavior of (\ref{Sigma(1)})
 as
\begin{equation}\label{ass1}
\frac B{B_{cr}}\rightarrow\infty.
\end{equation}
  The integrals in (\ref{Sigma(1)})
  do not contain
singularities in the integrands at $t=0$, but would diverge at
$t\rightarrow\infty$ if one just sets the limiting value
$\exp(-tB_{cr}/B)=1$ in the integrand. For this reason we should
divide the integration domain into two parts . Besides,
$(3g_2(t)/\sinh t)\rightarrow 1$ when $t\rightarrow\infty$ and,
hence, we have to add and subtract this limit beforehand in the
integrand of $\Sigma_2^{(1)}$. Handling the cases of $i=1,3$ does
not require the like, since $(g_{1,3}/\sinh t)\rightarrow 0$
sufficiently fast.
 \bea\label{Sigma(ass)}
\frac{3\pi}{2\alpha}\Sigma_i^{(1)}=
\int_0^\infty\exp\left(-\frac{tB_{cr}}{B}\right)
  \left(
  \frac{3g_i(t)}{\sinh t}-\frac
  1{t}-\delta_{i2}\right){\rm d}t
  +\delta_{i2}\int_0^\infty\exp\left(-\frac{tB_{cr}}{B}\right){\rm d}t
  \hspace{85mm}\nonumber\\[4pt]
  =\int_0^T\exp\left(-\frac{tB_{cr}}{B}\right)
  \left(
  \frac{3g_i(t)}{\sinh t}-\frac
  1{t}-\delta_{i2}\right){\rm d}t+\frac{B}{B_{cr}}\delta_{i2}\hspace{105mm}\nonumber\\[4pt]
  +\int_T^\infty\exp\left(-\frac{tB_{cr}}{B}\right)
  \left(
  \frac{3g_i(t)}{\sinh t}-
  \delta_{i2}\right){\rm d}t
  -\int_T^\infty\exp\left(-\frac{tB_{cr}}{B}\right)
\frac {{\rm d}t}{t} \hspace{105mm}
\nonumber\\[4pt]
=\int_0^T\left(
  \frac{3g_i(t)}{\sinh t}-\frac
  1{t}\right){\rm d}t-\delta_{i2}T
  +\int_T^\infty
  \left(
  \frac{3g_i(t)}{\sinh t}-
  \delta_{i2}\right){\rm d}t
   +\frac{B}{B_{cr}}\delta_{i2}\hspace{95mm}\nonumber\\[4pt]
+\ln\left(\frac{B_{cr}}{B}\right)+C+\ln T.
\hspace{55mm}(46)\hspace{85mm} \eea Here $T$ is an arbitrary
positive number, $\delta_{i2}$ is the Kronecker delta and $C$ is
the Euler constant. We have omitted the exponentials from the
first two integrals after the second equality sign in
(\ref{Sigma(ass)}), since now the resulting integrals converge,
and used the known asymptotic expansion of the standard
exponential-integral function, which is - up to terms linearly
decreasing with $B/B_{cr}$ - equal to\cite{abramovitz}
\begin{equation}\label{abram}
 -\int_T^\infty\exp\left(-\frac{tB_{cr}}{B}\right)
\frac {{\rm d}t}{t}=\left(\ln\frac{B_{cr}}{B}+\ln T+C\right)
\end{equation}

The most slowly decreasing term neglected from $3\int_T^\infty{\rm
d}t \exp\left(-tB_{cr}/B\right)
  g_i(t)/\sinh t$
  is $\delta_{i1}(3B_{cr}/4B)\ln (B_{cr}/B)$, since $g_1(t)/\sinh t\sim
  (1/4t^2)$ as $t\rightarrow\infty$. Other neglected terms
  decrease at least as fast as $B_{cr}/B$, since
  $(3g_i(t)/\sinh t)-\delta_{i2}$ decreases exponentially, like $\sim
  \exp(-2t)$, for $i=2,3$ when $t$ is large.

 Numerical calculations using MATHCAD code give for the
constants, involved in (\ref{Sigma(ass)}) (d$h_i/{\rm d}T=0$)
\begin{equation}\label{h}
  h_i=\int_0^T{\rm d}t\left(
  \frac{3g_i(t)}{\sinh t}-\frac
  1{t}\right)
  +\int_T^\infty\left(\frac{3g_i(t)}{\sinh t}-\delta_{i2}\right){\rm d}t+\ln
  T-\delta_{i2}T
\end{equation}
the following values:
\begin{equation}\label{constants}
  h_1=1.21,\hspace{5mm}h_2=-0.69,\hspace{5mm}h_3=0.21.
\end{equation}
Finally, in the asymptotic regime $B/B_{cr}\gg 1$, one has
\begin{equation}\label{fin}
  \Sigma_i^{(1)}=\frac{2\alpha}{3\pi}\left(\ln\frac{B_{cr}}{B}+C
  +h_i+\frac{B}{B_{cr}}~\delta_{i2}\right)
\end{equation}
 with
the accuracy to terms, decreasing at least as fast as integral
powers of the ratio $B_{cr}/B$ and to the slower term
$(\alpha/2\pi)(B_{cr}/B)\ln (B_{cr}/B)$, omitted from
$\Sigma_1^{(1)}$.

Turn now to $\Sigma_i^{(2)}$ Eq.~(\ref{B}). This depends upon the
three arguments, as indicated in (\ref{B}). We shall be interested
in the asymptotic domain, depicted by the condition (\ref{ass1})
and
\begin{equation}\label{ass2}
  \frac{2eB}{z_1}\rightarrow\infty.
\end{equation}
As for the ratio $2eB/z_2$ we shall keep it finite, whenever it
makes sense.

The asymptotic expansion of (\ref{sigma1}), (\ref{sigma2}),
(\ref{sigma3}) in powers of $\exp (-t)$ and $\exp (\eta t)$
produces an expansion of (\ref{B}) into a sum of contributions
coming from the thresholds (\ref{thresh}), the singular behavior
in the threshold points originating from the divergencies of the
t-integration in (\ref{B}) near $t=\infty$ (see
\cite{annphys,trudy,book} for detail). The leading terms in the
expansion of (\ref{sigma1}), (\ref{sigma2}), (\ref{sigma3}) at
$t\rightarrow\infty$ are \bea\label{sigass1}
  \left.\left(\frac{\sigma_1(t,\eta)}{\sinh
  t}\right)\right|_{~t\rightarrow\infty}=\frac{1-\eta}{2}~\exp(t (\eta-1)))
  ,
  \eea\bea\label{sigass2}
\left.\left(\frac{\sigma_2(t,\eta)}{\sinh
  t}\right)\right|_{~t\rightarrow\infty}=\frac{1-\eta^2}{4}~\left(1+2\exp(-2t)\right),
\eea
  \bea\label{sigass3}
  \left.\left(\frac{\sigma_3(t,\eta)}{\sinh
  t}\right)\right|_{~t\rightarrow\infty}=2\exp(-2t).
\eea With the change of the variable $\tau=t/eB$ and taking into
account that it follows from (\ref{M})  that $M(\infty,\eta)=1/2$,
one obtains for (\ref{B}) near the lowest singular thresholds
($n=0,$ $n'=1$ or  $n'=0,$ $n=1$ for $i=1$ in (\ref{thresh}),
$n=n'=1$ for $i=3$, and $n=n'=0$ and $n=n'=1$ for $i=2$)
\bea\label{Sig(2)i}
  \Sigma_i^{(2)}
  =
  \frac{2 \alpha eB}{\pi}\int_{-1}^1{\rm
  d}\eta\int_0^\infty{\rm d}\tau ~\exp(-m^2\tau)~\left.\left(\frac{\sigma_i(eB\tau,\eta)}{\sinh eB\tau}\right)\right|_{~t\rightarrow\infty}\times\nonumber\\[4pt]
  \times\left[\exp\left(-\frac{z_2}{2eB}-
  \frac{z_1(1-\eta^2)}{4}\tau\right)-1\right].\hspace{15mm}
   \eea
After the integration over $\tau$ we get, $e.g.$, for
    $\Sigma_1^{(2)}$
\bea \Sigma_1^{(2)}
  =
  \frac{4\alpha eB}{\pi}\int_{-1}^1{\rm
  d}\eta ~(1-\eta)\times\hspace{80mm}\nonumber\\[4pt]
\times\left(\frac
  {\exp\left(-\frac{z_2}{2eB}\right)}{4m^2+4(1-\eta )eB+z_1(1-\eta^2)}-
  \frac 1{4m^2+4(1-\eta )eB}\right).\hspace{20mm}
\nonumber
\eea
\begin{equation}\label{Sig(2)1}
\end{equation}
The pole in the above expression, caused by  the integration over
$t$, turns into the inverse square root singularity after the
integration over $\eta$ ($cf.$ derivation of (\ref{4}) from
(\ref{2})). In the limit (45), (51), when
$B\gg m^2$, $~B\gg|z_1|$, no singularity remains in this
expression (it is shifted to infinitely remote region) and we are
left with $\Sigma_1^{(2)}=(2\alpha/\pi)(\exp (-z_2/2eB)-1)$. The
same situation occurs for $\Sigma_3^{(2)}$, and for higher
thresholds (also for contributions into $\Sigma_2^{(2)}$ other,
than those coming from the first term in (\ref{sigass2})). The
result of the calculation analogous to (\ref{Sig(2)1}) for
$\Sigma_3^{(2)}$ is $(4\alpha/\pi)(\exp (-z_2/2eB)-1)$. We
conclude that in the limit (45), (51) there
are no cyclotron resonances in the eigenvalues $\kappa_{1,~3}$
according to (\ref{ro123}), also that $\Sigma_1^{(2)}$ does not
introduce a singular contribution into $\kappa_2$. Consequently,
there is no reason to keep the ratio $eB/z_2$ finite, when
$B\rightarrow\infty$, in $\Sigma_{1,~ 3}$, since $z_2$ may
infinitely grow on the dispersion curve only when there is a
resonance.

Therefore, we should consider only the limit, when all the three
arguments in $\Sigma_{1,3}^{(2)}$  tend to zero. Handling this
limit in (\ref{B}) is straightforward:
\begin{equation}\label{1/B}
  \lim \Sigma_{1,3}^{(2)}=\frac {-2\alpha}{\pi eB}\int_0^\infty {\rm
  d}t\int_{-1}^1{\rm d}\eta~\frac{\sigma_{1,3}(t,\eta)}{\sinh t}
  ~\left(z_2M(t,\eta)+z_1\frac{1-\eta^2}{4}t\right).
\end{equation}
The both integrations here converge, and hence this contribution
decreases like $z_1/eB$ and $z_2/eB$ when $B\rightarrow\infty$.
This is to be neglected within our scope of accuracy.

The situation is different with $\Sigma_2^{(2)}$. The resonance
behavior is present here due to the contribution of the
leading asymptotic term $(1-\eta^2)/4$ in (\ref{sigass2}). It is
responsible for the first threshold at $-z_1=4m^2$ ( the ground
Landau state $n=n'=0$ in (\ref{thresh})), which remains in its
place when $B\rightarrow\infty$. Hence we should keep the ratio
$z_2/eB$ nonzero when passing to the limit of large fields (since
$z_2\rightarrow\infty$ near the singular threshold on the
dispersion curve) for the contribution of this term into
$\Sigma_2^{(2)}$. The contributions of nonleading terms in the
expansion (\ref{sigass2}) into $\Sigma_2^{(2)}$ are nonsingular
and should be  treated along the same lines as $\Sigma_{1,3}$
above. They decrease like $~z_1/eB$, $~z_2/eB~$ to be neglected.
Finally, for (\ref{B}) we are left in the limit (\ref{ass1}),
(\ref{ass2}) with

\bea\label{lead}
  \Sigma_2^{(2)}=
  \frac{2\alpha}{\pi}\int_0^\infty{\rm d}t~
  \exp\left(-\frac{tB_{cr}}{B}\right)\int_{-1}^1{\rm
  d}\eta\frac{1-\eta^2}{4}\times\nonumber\\[4pt]
  \times\left[\exp\left(-z_2\frac{M(t,\eta)}{eB}-z_1
  \frac{1-\eta^2}{4eB}t\right)-1\right].
\eea By the change of the integration variable $t=eB\tau$ and the
use of the asymptotic form $M(eB\tau,\eta)=1/2$, $~eB\tau\gg1$ we
finally obtain - after the $\tau$-integration -  the leading
contribution into $\Sigma_2^{(2)}$ in the limit (\ref{ass1}),
(\ref{ass2}) \bea\label{Sigma(2)2}
 \Sigma_2^{(2)}
  = \frac{2\alpha eB}{\pi}\int_{-1}^1{\rm
  d}\eta ~(1-\eta^2)
\frac
  {\exp\left(-\frac{z_2}{2eB}\right)}{4m^2+z_1(1-\eta^2)}
  -\frac{2\alpha eB}{3\pi m^2}.\eea

 By combining Eqs.(\ref{Sigma(2)2}) and (\ref{fin}) according to
  (\ref{ro123}), and bearing in mind
(\ref{momenta}) and (\ref{Sigma})
  , we come to the expressions (\ref{1}), (\ref{2}), (\ref{kappa3})
  for the polarization operator eigenvalues, given in the body of
  the paper above, provided that we also neglect the constant and
  logarithmic terms in $\kappa_2$  coming from
  $\Sigma_{1,2}^{(1)}$ (\ref{fin}) as compared to the terms growing linearly
  with $B$.

\end{document}